\documentclass[conference,a4paper]{APSIPA2021}
\usepackage{cite}
\usepackage[numbers]{natbib}
\usepackage[colorlinks,citecolor=black]{hyperref}
\usepackage{multirow}
\usepackage{amsmath}
\usepackage[psamsfonts]{amssymb}
\usepackage{amsfonts}
\usepackage{amsxtra}
\usepackage{threeparttable}
\usepackage{graphicx}
\usepackage{bm}
\usepackage{geometry}
\usepackage{fancyhdr}

\fancypagestyle{firststyle} {
    \fancyhf{}
    \fancyhead[L]{Proceedings of 2022 APSIPA Annual Summit and Conference}
    \fancyhead[R]{7-10 November 2022, Chiang Mai, Thailand}
    
    \fancyfoot[L]{978-616-590-477-3 ©2022 APSIPA}

    \fancyfoot[R]{APSIPA ASC 2022}
}
\fancypagestyle{fancy} {
    \fancyhf{}
    \fancyhead[L]{Proceedings of 2022 APSIPA Annual Summit and Conference}
    \fancyhead[R]{7-10 November 2022, Chiang Mai, Thailand}
    
}

\begin{document}

\title{Speaker Representation Learning via Contrastive Loss with Maximal Speaker Separability}

\author{%
\authorblockN{%
Zhe LI and 
Man-Wai MAK
}
\authorblockA{%
The Hong Kong Polytechnic University, Hong Kong SAR\\
E-mail: lizhe.li@connect.polyu.hk, man.wai.mak@polyu.edu.hk
}
}
\maketitle

\begin{abstract}
A great challenge in speaker representation learning using deep models is to design learning objectives that can enhance the discrimination of unseen speakers under unseen domains. This work proposes a supervised contrastive learning objective to learn a speaker embedding space by effectively leveraging the label information in the training data. In such a space, utterance pairs spoken by the same or similar speakers will stay close, while utterance pairs spoken by different speakers will be far apart. For each training speaker, we perform random data augmentation on their utterances to form positive pairs, and utterances from different speakers form negative pairs. To maximize speaker separability in the embedding space, we incorporate the additive angular-margin loss into the contrastive learning objective. Experimental results on CN-Celeb show that this new learning objective can cause ECAPA-TDNN to find an embedding space that exhibits great speaker discrimination. The contrastive learning objective is easy to implement, and we provide PyTorch code at \url{https://github.com/shanmon110/AAMSupCon}.

\end{abstract}

\section{Introduction}
Speaker representation learning is a crucial process in speaker verification. Its objective is to learn a feature embedding space with the following attributes: 1) same-class compactness, where the embedding vectors of the same speaker are nearby; 2) different-class dispersion, where the embedding vectors belonging to different speakers are far apart. Because of the advances in deep
neural network (DNN) architectures \cite{snyder2017deep,snyder2018x,desplanques2020ecapa}, loss functions \cite{wan2018generalized,chung2020defence,wang2018cosface,deng2019arcface}, pooling methods \cite{cai2018exploring,okabe2006attentive}, and domain adaptation \cite{sang2020open,bhattacharya2019generative,sang2021deaan}, significant progress in speaker representation learning has been made in recent years. However, the models are still not sufficiently robust to noisy labels \cite{zhang2018generalized,sukhbaatar2015training} and are sensitive to input perturbation unless a notion of margin is introduced to their loss function \cite{elsayed2018large,liu2016large}. Studies have shown that these deficiencies can reduce generalization performance \cite{lin2018multisource,li2020contrastive,tu2020variational,tu2020information}.

\begin{figure}[ht] 
\includegraphics[width=0.5\textwidth]{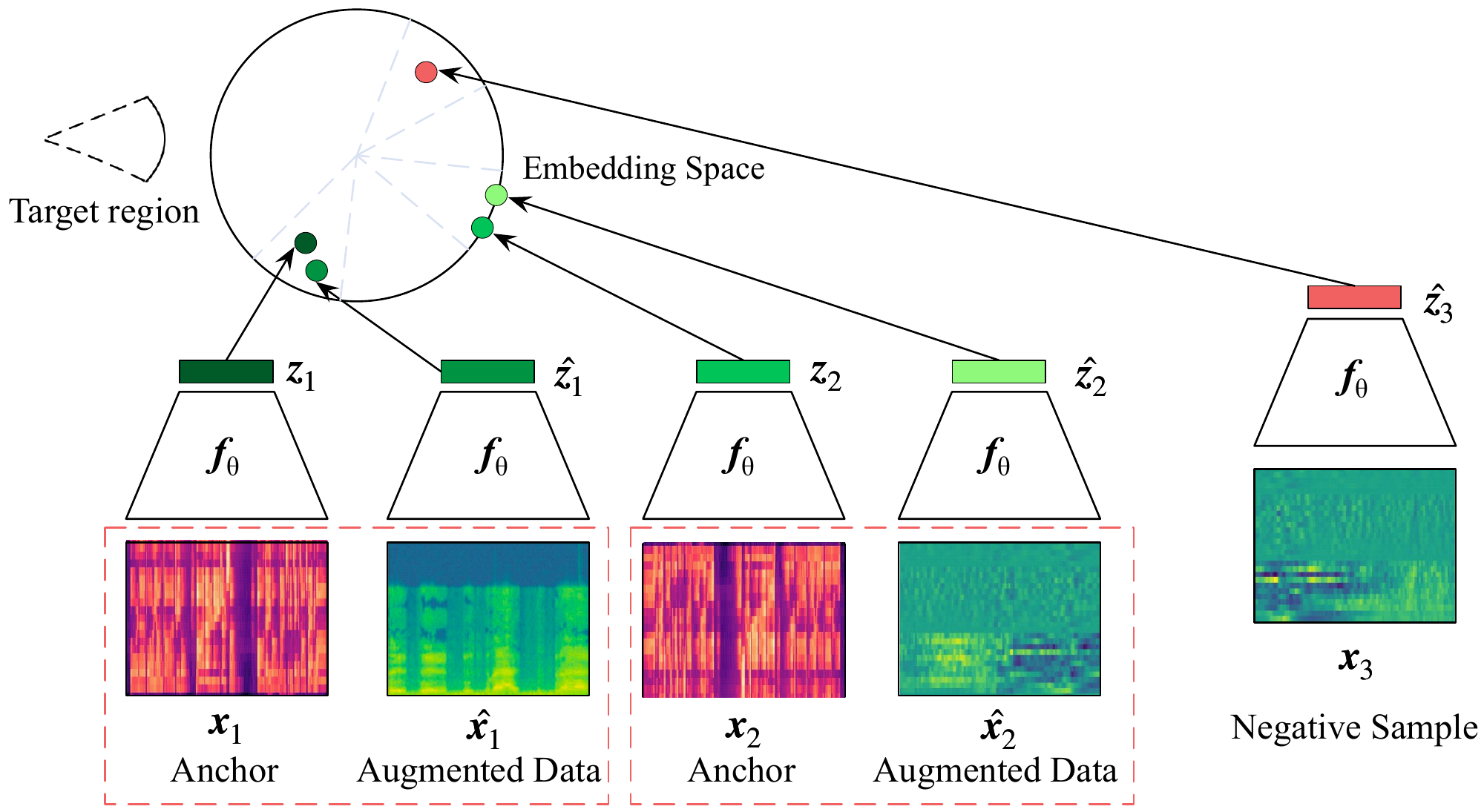}
\caption{Illustration of our basic idea. Learning an embedding space in which similar speakers pairs stay close to each other while dissimilar ones are far apart.}
\label{fig:basicidea}
\end{figure}

Several strategies have been developed to limit intra-class deviation and increase intra-class separation. For example, Wen {\it et al.} \cite{ wen2016discriminative} penalized the gaps between the features and their centers by adding a regularization term. The authors in \cite{ wang2017normface,liu2017rethinking,ranjan2017l2} proposed using a scale parameter to regulate the temperature \cite{ hinton2015distilling} of the softmax loss, causing well-separated samples to produce larger gradients, thereby shrinking intra-class dispersion. The authors in \cite{liu2016large} proposed enlarging the classification margin to make the learning objective harder, which encourages the learning of discriminative features. Liu {\it et al.} \cite{liu2017sphereface} proposed an angular distance metric in which the dissimilarity of objects is measured by their geodesic distance in a hypersphere manifold. They also introduced an angular margin in the distance measure to make the decisions more stringent. Liu {\it et al.} \cite{liu2017rethinking}, Liang {\it et al.} \cite{liang2017soft}, and Ranjan {\it et al.} \cite{ranjan2017l2} enhanced the softmax loss function by introducing various margins.

Significant advancements in self-supervised representation learning have been made recently because of the resurgence of contrastive learning \cite{wu2018unsupervised,henaff2020data,oord2018representation,tian2020contrastive,hjelm2018learning,chen2020simple,he2020momentum}. These works share the same concept: To learn an embedding space where positive pairs are near and negative ones are far apart. Due to the absence of labels, each positive pair often comprises an anchor and the augmented sample of the anchor, while a negative pair consists of randomly picked samples from the mini-batch except for the anchor. In \cite{oord2018representation,tian2020contrastive}, the relationship between contrastive loss and mutual information was discovered.

In this work, we propose \textbf{AAMSupCon}---a speaker representation learning method that uses \textbf{a}dditive \textbf{a}ngular \textbf{m}argin \textbf{sup}ervised learning and \textbf{con}trastive learning to leverage label information in training data. As shown in Fig.~\ref{fig:basicidea}, the embeddings of the same class are brought together, and those from different classes are pushed apart. To maximize speaker separability, we compute the angle between the weight vector of the ground truth class and the embedding using the arc-cosine of the logit and add a margin to the angle, which is followed by taking the cosine of the enlarged angle to recalculate the target logit \cite{deng2019arcface}. We investigated using multiple positive samples per anchor instead of a single positive in self-supervised contrastive learning. These positive samples are collected from samples of the same class as the anchor, as opposed to self-supervised learning in which positive samples come from the anchor's augmented data only.

Our main contribution is that we modified the contrastive loss function to utilize various positive samples per anchor and introduce an additive angular margin to the target class's angle in the embedding space, thereby extending contrastive learning to speaker verification with class boundaries robust to input perturbation. We demonstrated that the modified contrastive loss outperforms previous supervised learning in a speaker verification task.

\section{Methodology}
\begin{figure*}[htpb]
\centering
\begin{minipage}[t]{\linewidth}
\includegraphics[width=\textwidth]{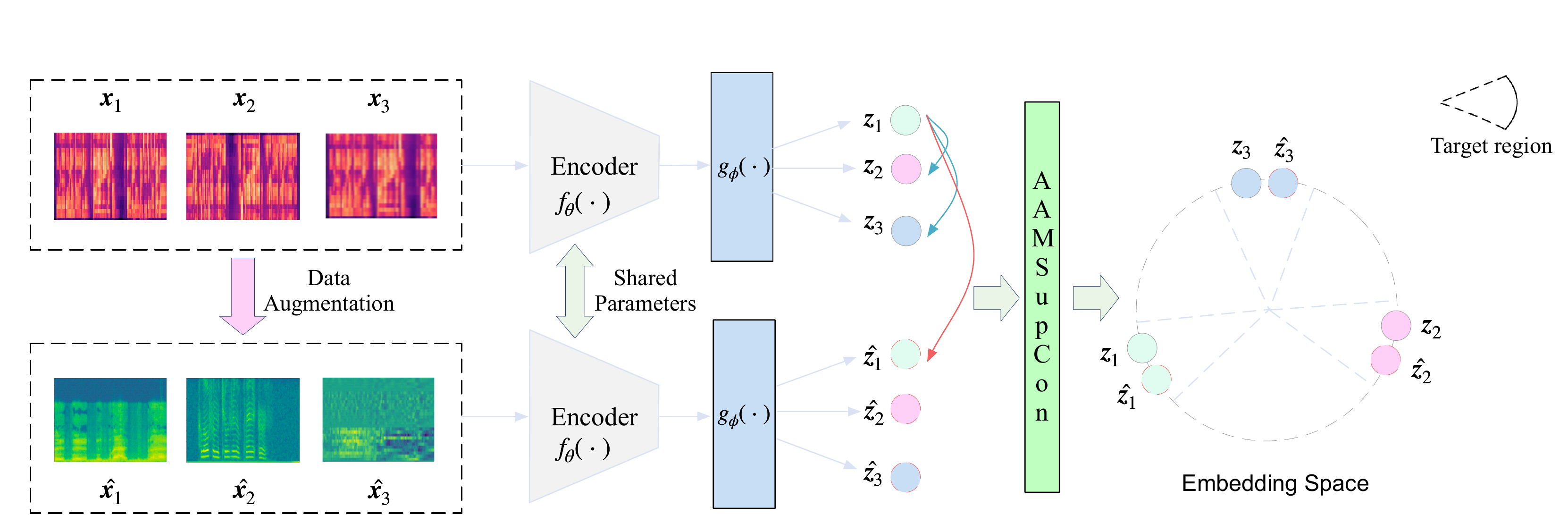}
\caption{The proposed architecture leverages additive angular margin loss and supervised contrastive learning. The encoder transforms the acoustic features (MFCC or FBank) to low-dimensional normalized embedding vectors. Invariance occurs for the embeddings (e.g., $\bm{z}_1$ and $\hat{\bm{z}}_1$) whose acoustic features ($\bm{x}_1$ and $\hat{\bm{x}}_1$) come from the same speaker. On the other hand, embeddings (e.g., $\bm{z}_1$ and $\bm{z}_2$) whose acoustic features ($\bm{x}_1$ and $\bm{x}_2$) belong to different speakers are far apart. The blue boxes represent the projection network. See text for details.}
\label{supcon}
\end{minipage}
\end{figure*}

Our objective is to train a feature embedding network using labeled audio. The embedding vectors of similar speakers should be close to each other, whereas those from different speakers should be far apart. Our method is fundamentally similar to the supervised contrastive learning described in \cite{khosla2020supervised,chen2020simple}, with adjustments for supervised classification. Given a batch of input samples, we perform data augmentation once to produce an augmented batch. The speaker characteristics of the embedding vector from the same instance should remain unchanged under various augmentations, while the embedding from different instances should be distinct. As shown in Fig.~\ref{supcon}, the spectrograms of both instances (the originals and their augmentations) are presented to an encoder network to produce 128-dimensional normalized embeddings. A projection network maps the embedding vectors to the output layer. At the outputs of the network, an additive angular margin contrastive loss is computed.

In this section, we will introduce our framework for representation learning and then examine the current supervised contrastive loss and additive angular margin loss. Then, we propose a margin contrastive loss function with remarkable discrimination. We conclude by comparing our framework with previous work.

\subsection{Representation Learning Framework}
Inspired by recent contrastive learning methods, AAMSupCon learns representations by maximizing the agreement across various augmented views of the same data through a contrastive loss in the latent space. As illustrated in Fig.~\ref{supcon}, this framework consists of four key components.

\paragraph{Data Augmentation} We produce one random augmentation for each input sample, $\hat{\bm{x}} = Augmentation(\bm{x})$. Each augmentation provides a unique perspective of the data and comprises a portion of the original sample's information. Using the Kaldi recipe \cite{snyder2019speaker}, we augmented the original utterances with noise, music, and chatter from the MUSAN dataset \cite{snyder2015musan}. We also generated reverberation effects by convolving the original waveforms with the RIR \cite{ko2017study} dataset's room impulse responses.

\paragraph{Encoder Network} Our objective is to train an encoder network $f_{\theta}(\cdot)$ from a set of labeled audio ${\cal X} = \{\bm{x}_1, \bm{x}_2,\ldots, \bm{x}_n\}$. $f_{\theta}(\cdot)$ transforms the input audio $\bm{x}_i$ to a low-dimensional embedding vector $\boldsymbol{h_i} = f_{\theta}\left(\bm{x}_{i}\right) \in \mathbb{R}^d$, where $d$ is the output dimension. Both the original and the augmented samples are independently fed to the same encoder, resulting in two representation vectors. For simplicity we chose ECAPA-TDNN \cite{desplanques2020ecapa} as the encoder.

\paragraph{Projection Network} It is a shallow network $g_\phi(\cdot)$ that transforms the encoder's output to a space in which contrastive loss is applied. $g_\phi(\cdot)$ is an MLP with one hidden layer and a linear output layer. We denote the transformed embedding vector as $\boldsymbol{z}_{i}=g_\phi\left(\boldsymbol{h}_{i}\right)= \bm{W}_2 \sigma\left(\bm{W}_1 \boldsymbol{h}_{i}\right)$, where $\sigma$ is a ReLU nonlinearity. We normalize the output of this network so that the embedding vectors lie on a unit hypersphere, which allows us to estimate the distance in the projection space using an inner product.

\subsection{Additive Angular Margin Contrastive Loss}
We explain how to incorporate additive angular margin into supervised contrastive learning. 

\subsubsection{\textbf{Supervised Contrastive Losses}}
\textbf{Sup}ervised \textbf{con}trastive loss (SupCon) can handle the situation where multiple samples are known to belong to the same class due to the presence of labels:
\begin{equation}
    L_{SupCon} = \sum_{i=1}^N \frac{-1}{\vert {\cal P}(i) \vert} \sum_{p \in {\cal P}(i)} \log \frac{\exp(\bm{z}_i \cdot \bm{z}_p /\tau)}{\sum_{a \in {\cal A}(i)}\exp(\bm{z}_i \cdot \bm{z}_a /\tau)} .
\label{Supcon}
\end{equation}
In Eq.~\ref{Supcon}, ${\cal P}(i)$ contains the indices of positive samples in the augmented batch (original + augmentation) with respect to $\bm{z}_i$ and $|{\cal P}(i)|$ is the cardinality of ${\cal P}(i)$. $\boldsymbol{z}_i$ is an anchor. $\boldsymbol{z}_a$ are negative samples. $\boldsymbol{z}_p$ are positive samples and ${\cal A}(i)$ is the index set of negative samples.

\subsubsection{\textbf{Additive Angular Margin Loss}}
Additive angular margin (ArcFace) loss calculates the angle between the embedding vector and the class weight vector using the arc-cosine function. It then adds an additive angular margin to the angle and uses the cosine function to recover the target logit. Then, ArcFace rescales all logits by a scale factor, and the remaining computations are identical to those of the softmax loss. Due to its precise relationship with geodesic distance on a hypersphere, the ArcFace has a clear geometric explanation. The ArcFace equation is:
\begin{equation}
    L_{ArcFace} = -\frac{1}{N} \sum_{i=1}^N \log \frac{e^{s(\cos(\theta_{y_i}+m))}}{e^{s(\cos(\theta_{y_i}+m))}+\sum_{j=1,j \neq y_i}^C e^{s\cos(\theta_j)}}
\label{ArcFace}
\end{equation}
where $\cos \theta_{y_i}$ is the target logit, which is the dot product of the normalized class-weight vector and the normalized embedding vector. $m$ is an additive angular margin that increases intra-class compactness and inter-class disparity. $C$ is the number of speakers.

\subsubsection{\textbf{Additive Angular Margin Supervised Contrastive Loss}}
\label{aamSupervised}
We propose an additive angular margin supervised contrastive softmax for supervised embedding learning by combining SupCon and ArcFace:
\begin{equation}
\begin{aligned}
L&_{AAMSupCon} = \\
&-\frac{1}{N} \sum_{i=1}^N \log \frac{e^{s(\cos(\theta_{y_i}+m))}}{e^{s(\cos(\theta_{y_i}+m))}+\sum_{j=1,j \neq y_i}^C e^{s\cos(\theta_j)}} \\ 
&+ \sum_{i=1}^N \frac{-1}{\vert {\cal P}(i) \vert} \sum_{p \in {\cal P}(i)} \log \frac{\exp(\bm{z}_i \cdot \bm{z}_p /\tau)}{\sum_{a \in {\cal A}(i)}\exp(\bm{z}_i \cdot \bm{z}_a /\tau)}   
\label{aamsupcon}
\end{aligned}
\end{equation}

AAMSupCon has the following advantages:
\begin{itemize}
\item Generalization to arbitrary positives. For each anchor in a multiview batch, its augmented sample and the other samples with the same label as the anchor contribute to the numerator of the loss function (Eq.~\ref{Supcon}). The supervised loss motivates the encoder to provide tightly matched representations for all entries of the same class, leading to a more compact speaker clusters in the embedding space.
\item Contrastive power improves with additional negatives. Eq.~\ref{aamsupcon} has a sum over the negatives in the denominator of the loss function. As a result, the capability to distinguish between noise and signal is enhanced when more negative samples are added. This trait is essential for representation learning, with several articles demonstrating impressive performance by increasing the number of negatives \cite{henaff2020data,he2020momentum,tian2020contrastive,chen2020simple}.
\item AAMSupCon optimizes the geodesic distance by the virtue of the perfect relationship between the angle and arc-cosing in the normalized hypersphere. Consequently, the AAMSupCon loss may ostensibly impose a more pronounced separation between the closest classes.
\end{itemize}
\subsection{Comparison with SupCon and ArcFace}
The AAMSupCon loss closely resembles the SupCon loss and the ArcFace loss. The SupCon loss is an innovative contrastive loss function that permits various positives per anchor. To maximize class separability, ArcFace adds an angular margin to the angle between the sample and the class weight. AAMSupCon combines the benefits of ArcFace and SupCon, exploits the label information for contrastive learning, and produces highly discriminative features for speaker verification.

\section{Experiments}
We evaluated the AAMSupCon loss on a speaker verification dataset called CN-Celeb \cite{fan2020cn,li2022cn}. For the encoder network, we experimented with the ECAPA-TDNN \cite{desplanques2020ecapa} architecture. The representation (embedding) vectors were extracted at the final pooling layer after normalization. We experimented with room impulse responses, music, background noise, and babble noise as four data augmentation types.
\begin{table}[ht]
\centering
\caption{Statistics of CN-Celeb 1 \& 2}
\setlength{\tabcolsep}{1.5mm}{
\begin{tabular}{r|rrrr}
\hline
Train/Test sets & Speakers & Recordings & Trials & Target trials\\
\hline
CNCeleb1 Train & 797 & 107,953 & N/A & N/A \\
CNCeleb2 Train & 1996 & 524,787 & N/A & N/A \\
CNCeleb Train (1\&2) & 2793 & 632,740 & N/A & N/A \\
CNCeleb1 Eval & 200 & 17973 & 3,484,292 & 17,755 \\
\hline
\end{tabular}
}
\label{cnceleb}
\end{table}
\subsection{Dataset}
We employed CN-Celeb \cite{fan2020cn,li2022cn}, which comprises CN-Celeb1 \cite{fan2020cn} and CN-Celeb2 \cite{li2022cn} to conduct our experiments. Table~\ref{cnceleb} lists the train-eval splits, speaker count, number of recordings, and evaluation trial statistics for the CN-Celeb dataset.
We utilized the training data in CN-Celeb1\&2, which include over 2,793 speakers with 11 genres to train our models. The genres include "advertisement", "drama", "entertainment", "interview", "live broadcast", "movie", "play", "recitation", "singing", "speech", and "vlog". Performance was evidenced using the evaluated set of CN-Celeb1.

\subsection{Evaluation Metrics}
We used the equal error rate (EER) and minimum detection cost function (minDCF, $\mbox{p-target}=0.01$) as the metrics to assess the performance.
\subsection{Experimental Setup}
We adopted 80-dimensional Fbank as input features. In the last step of the augmentation process, we utilized SpecAugment \cite{park2019specaugment} on the log-mel spectrograms. For each utterance, we randomly masked between 0 to 10 frames in the time domain and between 0 to 8 channels in the frequency domain. An SGD optimizer was used to train the models. We set the margin $m$ to 0.2. The mini-batch size for training is 3072. The contrastive learning temperature $\tau$ was set to 0.07.
\begin{table}[ht]
\caption{The experimental results of the proposed AAMSupCon and conventional methods on the CN-Celeb evaluation set. For each metric, the best and the second-best are highlighted in bold and underline, respectively.}
    \centering
    \begin{tabular}{r|crr}
\hline
Network & Loss Function & EER(\%) & minDCF \\
\hline
TDNN \cite{li2022cn} & Softmax & 12.39 & 0.60 \\
TDNN \cite{alam22b_odyssey}  & Softmax &  11.33 & 0.57  \\
ETDNN \cite{alam22b_odyssey} & Softmax &  11.09 & 0.56 \\
ETDNN-CA \cite{alam22b_odyssey} & Softmax & 10.88 & 0.56 \\
ETDNN-LSTM-CA\cite{alam22b_odyssey} & Softmax & 10.30 & 0.55 \\
HNN \cite{alam22b_odyssey} & Softmax &  9.18 & \underline{0.50} \\
MSHNN \cite{alam22b_odyssey} & Softmax & 9.05 & \textbf{0.48} \\
ENSEMBLE \cite{alam22b_odyssey} & Softmax   & \underline{8.94} & \textbf{0.48} \\
ResNet34 & Real AM-Softmax \cite{li2022real} & 11.05 & N/A \\
ECAPA-TDNN \cite{desplanques2020ecapa} & AAMSupCon (ours) & \textbf{8.49} & \underline{0.50} \\
\hline
\end{tabular}
\label{mainresult}
\end{table}
\subsection{Results and Analysis}
Our system achieves the best performance, outperforming the second-best with the TDNN architecture by 0.43\% in terms of EER. The results in Table~\ref{mainresult} indicate that AAMSupCon is superior to other conventional systems on CN-Celeb. This suggests that the proposed loss function force the network to learn speaker features while maximizing discrimination. We conjecture that the good performance is due to (1) the capability of the proposed contrastive loss (Eq.~\ref{aamsupcon}) in capturing the correlation between samples from the same speaker and contrasting the samples from different speakers and (2) the angular margin that increases the tolerance to feature perturbation.

\begin{table}[ht]
\caption{The influence of loss functions on EER and minDCF. For each metric, the best and the second-best are highlighted in bold and underline, respectively.}
    \centering
\begin{tabular}{c|rrr}
\hline
Encoder&Loss Function&EER(\%) &minDCF \\
\hline
ECAPA-TDNN \cite{desplanques2020ecapa} & Softmax & 16.07 & 0.82 \\
ECAPA-TDNN \cite{desplanques2020ecapa} & AMSoftmax & 13.39 & \underline{0.71}  \\
ECAPA-TDNN \cite{desplanques2020ecapa} & RAMSoftmax & 13.25 & 0.72 \\
ECAPA-TDNN \cite{desplanques2020ecapa} & AAMSoftmax & \underline{8.79} & \textbf{0.50}  \\
ECAPA-TDNN \cite{desplanques2020ecapa} & AAMSupCon & \textbf{8.49} & \textbf{0.50} \\ 
\hline
\end{tabular}
\label{ablation}
\end{table}

\subsection{Ablation Study}
We conducted an ablation experiment to determine the contributions of critical components in our method. Precisely, we ran experiments with the same encoder but used different losses. Table~\ref{ablation} shows that Softmax achieves the worst performance because Softmax does not have margin. AM-Softmax achieves similar performance as RAM-Softmax, which has also been verified in \cite{li2022real}. Both AM-Softmax and RAM-Softmax have been widely employed in metric learning, aiming to maximize the difference between target pairs and non-target pairs. AAMSoftmax achieves the second best performance due to the use of additive angular margin in the loss functions, which enables it to obtain highly discriminative features. AAMSupCon achieves the best performance because it effectively leverages the label information so that samples of the same class can be packed closely. Moreover, we have added max-margin to AAMSupCon to increase intra-class attraction and inter-class repulsion, causing the target and non-target classes to be well separated.
\begin{table}[ht]
\caption{The influence of Batch size on EER and minDCF.}
    \centering
    \begin{tabular}{c|cc}
\hline
Batch Size & EER(\%) & minDCF \\
\hline
128  &  13.64 & 0.71  \\
512  &  11.03 & 0.66 \\
1024 &  10.27 & 0.65 \\
\hline
\end{tabular}
\label{batch}
\end{table}
\subsection{Batch Size}
Contrastive learning was carried out for each min-batch. As described in Section~\ref{aamSupervised}, the numbers of positives and negatives in each min-batch are critical to contrastive learning. More samples in a mini-batch can enable the network to learn more robust features. To verify our conjecture, we only used supervised contrastive loss and ECAPA-TDNN in the encoder. We set three different batch sizes, 128, 512 and 1024, as shown in Table~\ref{batch}. Apparently, a larger batch size achieves better performance. The experimental results verify our hypothesis.

\section{Conclusion}
This study proposes an additive angular margin supervised contrastive representation learning framework that successfully increases the discriminative capability of feature embeddings for speaker verification. This is achieved by using label information in contrastive learning. Experiments reveal that our technique routinely outperforms the baselines. To enable the repeatability of the findings described, codes and explanations are available on our GitHub site.

\section{Acknowledgment}
This work was in part supported by Research Grants Council of Hong Kong, Theme-based Research Scheme (Ref.: T45-407/19-N).

\bibliographystyle{IEEEtran}
\bibliography{ref}

\end{document}